\documentclass[letter]{aa}

\usepackage{graphicx}
\usepackage{txfonts}
\begin{document}

%\nolinenumbers

   \title{The star grinder in the Galactic centre}

   \subtitle{Uncovering the highly compact central stellar-mass black hole cluster}

   \author{J. Haas\inst{1}\thanks{\email{haas@sirrah.troja.mff.cuni.cz}},
           P. Kroupa\inst{1,2}, L. \v{S}ubr\inst{1}
           \and M. Singhal\inst{1}}

   \institute{Charles University, Faculty of Mathematics and Physics, Astronomical Institute,
              V Hole\v{s}ovi\v{c}k\'ach 2, CZ-18000 Prague, Czech Republic
              \and
              Helmholtz-Institut f\"{u}r Strahlen- und Kernphysik, University of Bonn,
              Nussallee 14-16, D-53115 Bonn, Germany
             }

% \abstract{}{}{}{}{}
% 5 {} token are mandatory
 
  \abstract
   {Various past theoretical considerations and observational efforts suggest the presence of
   a population of stellar-mass black holes in the innermost parsec of the Galactic centre.}
   {In this Letter, we investigate the impact of these black holes on the composition of
   the embedding stellar population through their direct collisions with the individual stars.
   Based on the estimated collision rates, we derive an order of magnitude radial density
   profile of the black hole cluster.}
   {The estimates were obtained analytically, considering various possible formation channels
   for the black holes and the observed present-day properties of the stellar populations
   in the Galactic centre.}
   {We find that the collisions of the stars and the black holes can lead to the depletion
   of the most massive stars within
   the S-cluster on a timescale of a few million years. The necessary black hole cluster density
   is compatible with the recurrent in situ star formation in the innermost parsec of the
   Galactic centre.
   We suggest that such a depletion naturally explains the reported lack of
   stars of the stellar type O and of the Galactic halo hyper-velocity star counterparts within
   the S-cluster.}
   {}

   \keywords{stars: black holes -- stars: kinematics and dynamics -- Galaxy: center}

   \maketitle
%
%--------------------------------------------------------------------------------------------------
%
\section{Introduction}
\label{intro}
The innermost parsec of the Galactic centre is generally believed to host a population of
stellar-mass black holes \citep[e.g.][]{Morris93, Mouawad05}.
Its structure and properties, however, remain largely unknown despite substantial theoretical
\citep[e.g.][]{Miralda-Escude00, Freitag06, Hopman06, Alexander09, Merritt10, Preto10,
Antonini12, Vasiliev17, Generozov18, Zhang24} as well as
observational \citep[e.g.][]{Hailey18, Mori21, Zhao22} efforts.
This is primarily due to the fact that observations of stellar-mass black holes typically rely on
the detection of radiation coming from the accreting surrounding material that may not always
be available. For
this reason, an unknown (and potentially large) fraction of the black hole population may remain
undetected. Consequently, observations may map, instead of the black hole population,
the distribution
of gas in the surveyed region. Recent work of \citet{Zhao22} reports a few tens of
`hyper-compact radio sources' that can represent
candidate massive stellar remnants in the central parsec of our Galaxy. The distribution of these
sources on the sky closely matches the morphology of the gas streamers known as the minispiral
\citep[e.g.][]{Nitschai20}.

Various possible and non-mutually exclusive sources of the stellar-mass black holes in the
Galactic centre have been suggested, each leading to different properties of the
resulting black hole population. The black holes
may have been born in the larger-scale \citep[a few pc in radius; e.g.][]{Schoedel14}
nuclear star cluster and accumulated in its central
parts dynamically. It has been shown by \citet{Baumgardt18} by means of numerical $N$-body
modelling, however, that during a Hubble-time-long evolution of the nuclear star cluster with
repetitive star formation events, only about 300 stellar-mass black holes migrate into the
innermost 0.1~pc of the Galactic centre. Hence, despite its great total mass, the nuclear star
cluster does not seem to be a very effective source of stellar-mass black holes for its
innermost region.

Some of the stellar-mass black holes could be relics from the formation of the
supermassive black hole Sgr~A$^\star$ itself in the model of \citet{Kroupa20}. In this model,
at the beginning of the formation of a later elliptical galaxy or bulge, a hypermassive star
burst cluster of quasar luminosity \citep{Jerabkova17} forms at its centre.
After the massive stars have died, the more than $10^5$ stellar-mass black holes
merge rapidly due to the
infall of gas from the still-forming elliptical or bulge, leading to the formation of the
supermassive
black hole within a few hundred million years, but also most likely leaving a remnant population of
stellar-mass black holes in its vicinity. At the moment,
however, no literature on the properties of the remnant black holes exists, to our knowledge.

Another way to accumulate stellar-mass black holes in the innermost parts of the
Galactic centre is the in situ formation of massive stars in gaseous accretion discs around
Sgr~A$^\star$.
This is a favourable formation scenario for the young ($\approx5$~Myr) stellar cluster observed
between 0.04~pc and 0.5~pc from Sgr~A$^\star$ in projection
\citep[e.g.][]{Levin03, Paumard06, Bartko10, Lu13, Yelda14, Fellenberg22}.
This cluster contains more than a hundred very massive OB and Wolf-Rayet stars that will
end their lives as black holes. Its overall present-day mass function is top-heavy, presumably
due to its formation in the extreme environment that directly yielded such a (perhaps also
lower-mass truncated) mass distribution
\citep[e.g.][]{Morris93, Levin03, Nayakshin07, Schoedel20, Kroupa24}.

In this Letter, we suggest that repetitive massive star formation in accretion discs
around Sgr~A$^\star$ may have given birth to a stellar-mass black hole cluster dense enough
to significantly alter the stellar-type abundances below the inner edge of the young
stellar cluster through direct collisions and grazing encounters of the individual black holes
and stars. Based on this premise, we derive an order-of-magnitude radial density profile for the
black hole cluster.
%
%--------------------------------------------------------------------------------------------------
%
\section{The grinder}
\label{grinder}
Star formation in the innermost parsec of our Galaxy is still a matter of an ongoing debate but
it was suggested to be recurrent \citep[e.g.][]{Morris96}. This is based on the notion
of a continuous inflow
of gas from farther regions that accumulates in the vicinity of Sgr~A$^\star$ and forms massive
stars whose winds prevent further gas inflow until they cease due to stellar evolution and
the whole cycle repeats itself.
If we take the currently observed cluster of young stars as representative of the star formation
in the innermost parsec of our Galaxy and assume that such stars have been forming there
continuously over cosmological
timescales, we can make the following estimate of the resulting stellar-mass black hole number
density.

The typical lifetime of a star that ends its life as a black hole can be roughly
estimated as 5~Myr. Hence, considering the roughly 100 such stars in the observed young cluster, we arrive at a star formation rate of
$2\times10^{-5}$~yr$^{-1}$, which yields about $2\times10^{5}$ black holes over 10~Gyr. Since
the young stars are located mostly within 0.1~pc from Sgr~A$^\star$, the corresponding
black hole number density is roughly $2\times10^{8}$~pc$^{-3}$.

The existence of the black hole cluster of the above density would, however, have
a significant impact on the stellar population through
collisions and close encounters of the stars with the
stellar-mass black holes. In order to quantify this, we assume
the number density, $n_\bullet $, of the black holes to be constant
within the volume of space where the interactions should be occurring. We define a direct collision
of the black hole and the subject star as an encounter with the closest approach distance,
$r_\mathrm{per}<R_\star$, with $R_\star$ being the physical radius of the star.
The rate of such collisions can be written as \citep{Binney08}
\begin{equation}
\label{dens}
1/t_\mathrm{coll} = 4\sqrt{\pi}n_\bullet\sigma R_\star^2\left(1+v_\star^2/2\sigma^2\right)\,,
\end{equation}
with $t_\mathrm{coll}$ being the time between collisions, $\sigma$ the velocity dispersion within
the black hole cluster, and $v_\star=\sqrt{\frac{2GM_\star}{R_\star}}$
the escape speed from the surface of the subject star of mass $M_\star$; $G$ stands for
the gravitational constant.

For simplicity, we assume that these direct collisions lead to destruction of
the impacted stars, regardless of the approach velocity. While this assumption is supported by the
hydrodynamical models of \citet{Kremer22} for slow (parabolic) approaches, detailed modelling
of faster (hyperbolic) cases is not available at the moment. A population of stellar-mass black
holes surrounding Sgr~A$^\star$ thus acts like a `star grinder', with any new star being
destroyed by collisions with the black holes on the timescale of
$t_\mathrm{coll}$.

\begin{figure}
\centering
\includegraphics[width=\columnwidth]{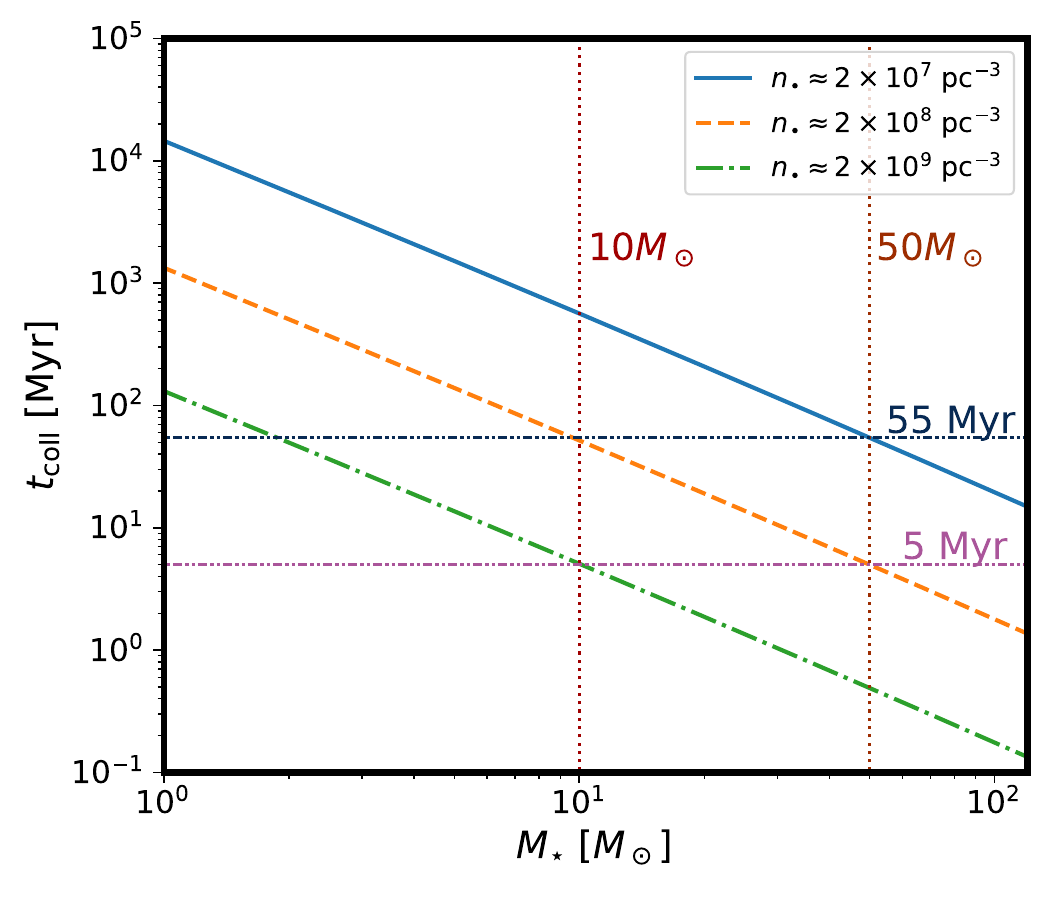}
\caption{Collision time, $t_\mathrm{coll}$, as a function of stellar mass, $M_\star$
(Eq.~(\ref{dens})), for three
different values of the black hole number density, $n_\bullet$. The horizontal dot-dot-dashed purple and dark blue lines denote the collision times of 5~Myr and 55~Myr for the
representative O and B stars of masses 50$\,M_\odot$ and 10$\,M_\odot$ (indicated by the vertical
dotted brick-red lines), respectively, derived for $n_\bullet=2\times10^8$~pc$^{-3}$.
In all cases, the velocity dispersion is set to the orbital (circular)
velocity around Sgr~A$^\star$ at the representative distance of 0.015~pc from Sgr~A$^\star$
and the mass-radius relation of \citet{Demircan91}, $R_\star=0.85\,M_\star^{0.67}$, is assumed.}
\label{tcoll}
\end{figure}
%
%--------------------------------------------------------------------------------------------------
%
\section{The ground-down S-cluster and beyond}
\label{s-cluster}
The star grinding effect is of special relevance for the roughly isotropic cluster
of young stars observed closer than 0.04~pc to Sgr~A$^\star$, the so-called S-cluster
\citep{Ghez03, Genzel03, Ghez05, Paumard06, Bartko10, Lu13, Yelda14, Habibi17, Fellenberg22}.
Its formation scenario is still debated but the S-cluster seems to be equally as old as the
young stars beyond 0.04~pc from Sgr~A$^\star$ \citep{Habibi17}.

The two young stellar structures,
however, differ dramatically in the abundance of the most massive stars.
While observations of the S-cluster revealed about two dozen B stars, no star of spectral type
O was found there. This is in striking contrast with the observations of the directly
neighbouring cluster of young stars at distances larger than 0.04~pc from Sgr~A$^\star$
where the numbers of the observed O and B stars are similar \citep[about 100;][]{Fellenberg22}.
Although different star formation conditions
may be invoked, it is tempting to estimate the effect of `the grinder'
given the astrophysical conditions within the S-cluster, which may relax the need for different
initial mass functions in the two neighbouring regions.

Assuming there was a population of O stars also initially present within the S-cluster but
entirely destroyed by the direct collisions with the stellar-mass black holes within
the age ($\approx5$~Myr) of their still-living supposed siblings located farther away,
the necessary black hole number density, $n_\bullet$, can be estimated as follows. We set
$t_\mathrm{coll}\approx5$~Myr and $\sigma\approx1100$~km/s equalling the orbital (circular)
velocity around Sgr~A$^\star$ at a representative distance of 0.015~pc from Sgr~A$^\star$.
For a representative O star, we set its mass to be $M_\star^\mathrm{O}=50\,M_\odot$, where
$M_\odot$ stands for the mass of the Sun.
According to the mass-radius relation of \citet{Demircan91}, $R_\star=0.85\,M_\star^{0.67}$,
we get $R^\mathrm{O}_\star\approx12\,R_\odot$, where $R_\odot$ denotes the Solar radius, and
the escape speed thus is $v^\mathrm{O}_\star\approx1260$~km/s.
The resulting black hole number density given by
Eq.~(\ref{dens}) is then $n_\bullet\approx2\times10^8~\mathrm{pc}^{-3}$. Remarkably, this value
is in agreement with the black hole number density estimate based on the star formation
arguments given in Sect.~\ref{grinder}.

If we now assume the obtained $n_\bullet$, we can estimate the destruction time for stars of any 
other spectral type and, specifically, for the stars of the spectral type B that are present
within the S-cluster. For the representative B star, $M^\mathrm{B}_\star=10\,M_\odot$,
$R^\mathrm{B}_\star\approx4\,R_\odot$, and $v^\mathrm{B}_\star\approx970$~km/s. Consequently,
$1/t_\mathrm{coll}\approx1/55~\mathrm{Myr}^{-1}$. Hence, while
the O stars within the S-cluster are destroyed in about 5~Myr, it takes 55~Myr to
shatter the B stars. As a result, the about 5~Myr old B stars in the S-cluster can still be
observed (see Fig.~\ref{tcoll}).
%
%--------------------------------------------------------------------------------------------------
%
\subsection{Radial density profile of the black hole cluster}
\label{profile}
If we assume that the initial mass function of the S-cluster and the young star cluster
above 0.04~pc from Sgr~A$^\star$ were at least similar at their high-mass ends, we
can use the currently observed occurrence of stellar types to constrain the
radial density profile of the stellar-mass black hole cluster.
In doing so, we implicitly assume that the distribution of the stellar types of the massive young
stars has been directly sculpted by the black holes. We note that this is where our approach
diverges from the usual considerations \citep[e.g.][]{Dale09, Davies11} in which
some radial density profile of the black hole cluster is assumed to investigate the impact of
the black holes on the environment.
\begin{figure}
\centering
\includegraphics[width=\columnwidth]{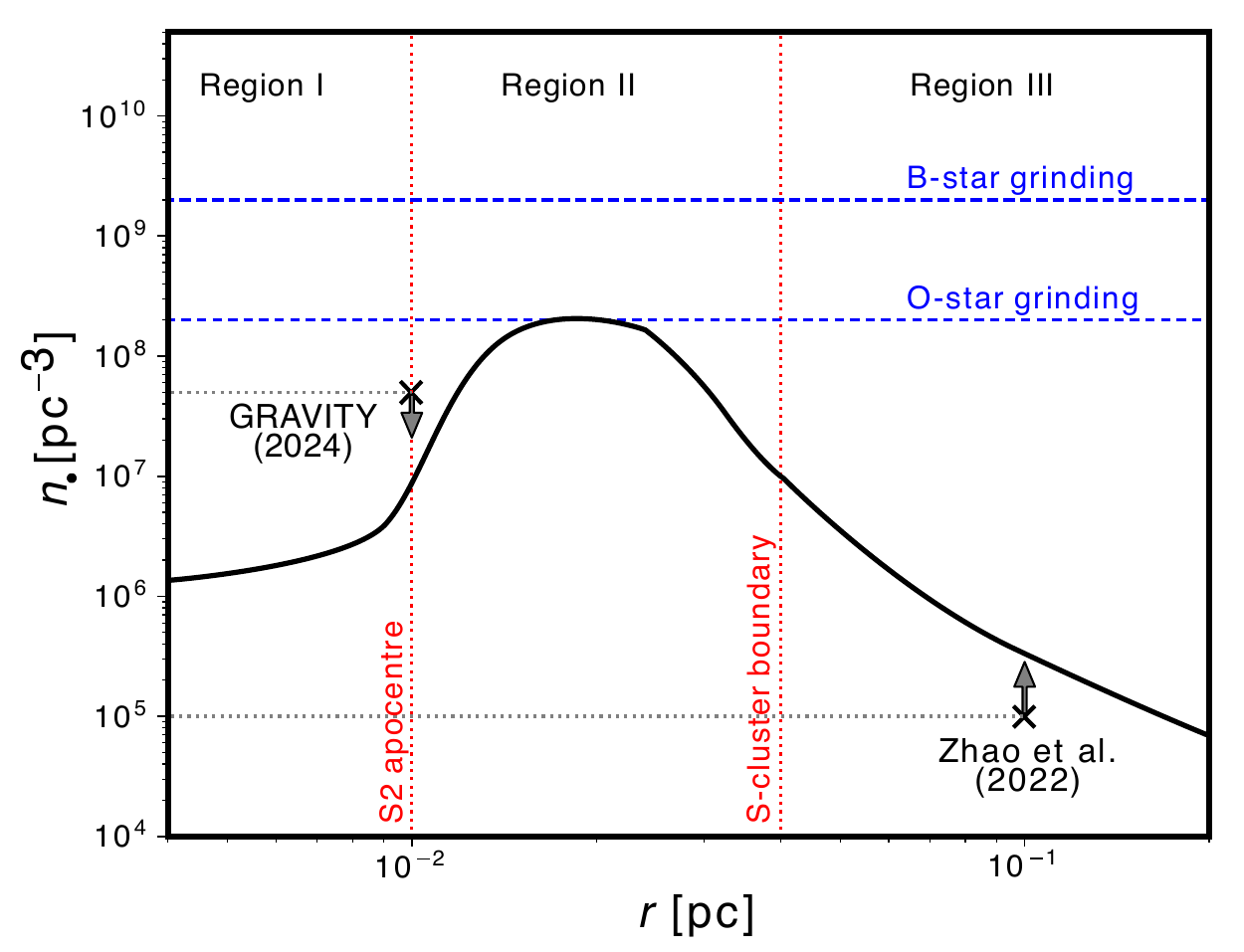}
\caption{Suggested constraints on radial number density profile, $n_\bullet\left(r\right)$,
of the stellar-mass black hole cluster as a function of radial distance, $r$, from the
Sgr~A$^\star$ supermassive black hole are plotted together with an arbitrary distribution function
(solid `bumpy' black line) fulfilling these constraints.
The vertical dotted red lines mark the boundaries
between regions I, II, and III in which the impact of the black holes is qualitatively different.
The horizontal dashed lines mark the density of the black holes for which O and B stars would be
ground down in 5~Myr. The observational constraints based on the \citet{GRAVITY24} enclosed mass
limits and on a subsample of the hypercompact radio sources from \citet{Zhao22} are
depicted by the x-shaped points with arrows at the corresponding distances from Sgr~A$^\star$.
The purpose of the dotted grey lines is to guide the eye of the reader to the particular density
values.}
\label{prof}
\end{figure}

In particular, since stars of stellar type B are observed within the S-cluster,
$n_\bullet$ cannot reach values $\gtrsim2\times10^9$~pc$^{-3}$ there.
Such a high black hole number density would lead to the collisional destruction of the B stars
within their estimated age of 5~Myr (see Fig.~\ref{tcoll}).
While this rather generous upper limit applies for the S-cluster as a whole, a
stronger constraint has recently been placed on the innermost part of the S-cluster.
Analyses of multiple stellar orbits there \citep{GRAVITY22, GRAVITY24} generally
suggest that no more than about $5\times10^3\,M_\odot$
(3$\sigma$ limit) of extended mass is present within the orbital apocentre
of the S2 star; that is, below 0.01~pc. Depending on the extended mass radial density profile
assumed, this upper limit may be even smaller by up to a factor of $\approx5$ \citep{GRAVITY24}.
In the case of the constant black hole number density, the derived constraint
for that region reads $n_\bullet\lesssim5\times10^7$~pc$^{-3}$ (see Fig.~\ref{prof}, region I).

This latter constraint implies that most of the `grinding' black holes need to be located beyond
the orbit of the S2 star but still within the S-cluster; that is, roughly between 0.01~pc and 0.04~pc
from Sgr~A$^\star$ (see Fig.~\ref{prof}, region II). In this region, their density should thus
be the highest, roughly reaching the grinding value of $n_\bullet\approx2\times10^8$~pc$^{-3}$.

Beyond the outer boundary of the S-cluster, at distances $\gtrsim0.04$~pc from
Sgr~A$^\star$ (see Fig.~\ref{prof}, region III), stars of stellar type O are abundant,
suggesting that $n_\bullet<2\times10^8~\mathrm{pc}^{-3}$ there again.
This constraint for the outer regions is in agreement
with the observations of the hyper-compact radio sources reported in \citet{Zhao22}. According to
their Fig.~1, the sources are most abundant around the IRS~13E complex, where about
10 of those labelled as candidate massive stellar remnants
are located within a circle of roughly 0.03~pc in radius. The corresponding number density
within a sphere of such a radius is about $10^5$~pc$^{-3}$.

Since the primary source of the stellar-mass black holes in our `grinder' model is the
recurrent massive star formation in (presumably similar) accretion discs around Sgr~A$^\star$,
the resulting black hole cluster is likely to be compact. The black hole number density,
$n_\bullet$, is thus probably steeply decreasing
over a few times 0.1~pc from Sgr~A$^\star$ down to the values
given by the stellar-dynamical evolution of the background larger-scale nuclear star
cluster.
These are negligible from the perspective taken here, as they are necessarily smaller than
about $3\times10^4$~pc$^{-3}$, a value obtained for the 300 black holes found
within 0.1~pc from Sgr~A$^\star$ according to the paper of \citet{Baumgardt18}.

The above-sketched radial density profile of the black hole cluster exhibits a maximum
just below or near the outer boundary of the S-cluster (see
the solid black curve in Fig.~\ref{prof}, region II). Such
a bump-like shape was predicted to be a possible result of the dynamical evolution of the
stellar-mass black hole population against a shallow-cusp stellar background
\citep{Merritt10}. A shallow-cusp radial density profile of the old stellar population has
indeed been observationally confirmed recently down to about 0.02~pc from Sgr~A$^\star$
\citep{Habibi19}. Furthermore, the hyper-compact radio sources (representing candidate massive
stellar remnants) reported in \citet{Zhao22}
also show a bump-like profile. Hence, even though the theoretical modelling is sensitive to the
choice of the underlying model
\citep[cf. e.g.][who did not find any bump for their set-up]{Generozov18}
and the observational detections of massive stellar remnants are likely not complete,
the qualitative
agreement with our independent results represents a strong supportive argument for the existence
of the stellar-mass black hole bump in the Galactic centre.
%
%--------------------------------------------------------------------------------------------------
%
\subsection{Hypervelocity-star counterparts}
\label{hyper}
Observations of the Galactic halo have revealed about two dozen stars moving with
velocities higher than the escape speed from the Galaxy, so-called
hypervelocity stars, some of which seem to have flight
directions compatible with a Galactocentric origin \citep{Brown14, Brown18}. One of the
explanations for the existence of such hypervelocity stars is the Hills mechanism
\citep{Hills88} occurring in the vicinity of Sgr~A$^\star$, which
is the tidal disruption of a binary star close enough to the supermassive black hole. Upon such a
disruption, one of the original binary components is left on a tight orbit around the
supermassive black hole, while the other is ejected with a
high velocity away from it.

The observed hypervelocity stars are of spectral type B
and are found at distances of about 50 kpc from the Galactic centre. Assuming an
average velocity of about 1000~km/s, these stars thus needed about 50~Myr to get
to their observed positions, which provides a lower estimate of their ages. Hence,
their equally old counterparts of the same spectral type (assuming a realistic
pairing in the original binaries) left in the S-cluster should still be observed there.
Although a detailed stellar-age survey of the S-cluster still needs to be
done, the B stars analyzed so far within this cluster are all younger than 25~Myr \citep{Habibi17}.

Based on the grinding timescale of 55~Myr derived for B stars in this Letter, we suggest that
while the young B stars can still be observed within the S-cluster, the older hypervelocity-star
counterparts were already destroyed by the collisions with the stellar-mass black holes.
%
%--------------------------------------------------------------------------------------------------
%
\section{Discussion}
\label{discuss}
Using two independent methods (continuous massive star formation and
their collisional grinding), we have obtained the same estimate of the stellar-mass
black hole number density, $n_\bullet\approx2\times10^8~\mathrm{pc}^{-3}$, within the S-cluster.
This value is rather high and it may be considered an upper
limit.

If we assumed such a density for the region
between 0.015~pc and 0.025~pc from Sgr~A$^\star$, the total number of the black holes there
would be about $10^4$. If we further set the mass, $m_\bullet$, of each individual black
hole to be $m_\bullet=10\,M_\odot$, the corresponding total mass of the black-hole component
of the S-cluster would be $10^5\,M_\odot$.
We note that these numbers are very rough estimates based on a simplified model of constant
$n_\bullet$ within a representative volume such that the
black holes orbitally cover the outer part of the S-cluster (see Fig.~\ref{prof}, region II).
The true values are dependent on the particular overall shape
and dimensions, density profile, and internal dynamics of the black hole cluster, which are
at the moment unknown. At the same time, it is not clear how to extrapolate the black hole cluster
radial density profile to the regions where the currently available observational constraints
directly apply \citep{Boehle16, GRAVITY22, GRAVITY24}. For this purpose, more detailed
investigations of the black hole cluster properties beyond the scope of this Letter are necessary.

The demands on the black hole number density, $n_\bullet$, are
somewhat decreased if we consider that
in addition to the direct collisions, grazing encounters of stars and the black holes can occur.
Such encounters are more frequent than the direct collisions due to their larger
cross-section and they can cause
a significant tidal stripping of stellar upper layers. The resulting mass loss
effectively changes O stars to stars of spectral type B, which enhances the disproportion of
O relative to B stars within the S-cluster. Owing to the very short Kelvin-Helmholtz
timescale of a B star of about $10^5$~yr, the affected stars would likely be observable as
normal main-sequence stars. Multiple subsequent grazing encounters may also eventually lead
to a complete destruction of the original O star.

If we define a grazing encounter of the black hole and the subject star as an encounter with
the closest approach distance, $r_\mathrm{per}$, such that the star entirely fills its Roche
lobe, we obtain $r_\mathrm{per}\approx1.6\,R_\star$ for a pair consisting of a 50~$M_\odot$ O star and a
10~$M_\odot$ black hole. The corresponding cross-section will thus be about a factor of 2.6
larger, reducing the grinding black hole number density, $n_\bullet$, by up to the same factor
(multiple grazing encounters may be necessary for a significant mass loss).
We note that smaller required values of $n_\bullet$ due to the grazing encounters
also lower the demands on the necessary black hole formation rate.

A very important aspect of the model introduced here that relies on close interactions
of the stars and the black holes (and thus requires a large value of $n_\bullet$) is the temporal
evolution of the black hole cluster density profile. The course of this evolution, however, not
only depends on the initial set-up of the black hole cluster (as well as the whole larger-scale
embedding stellar cluster) but can also be affected by further components of the system
that are not considered in this Letter. In particular, the recurrently infalling gas would,
in addition to fuelling the repetitive star (and thus black hole) formation, also lead to
dissipative interactions, and thus shrinkage of the orbits of the black holes. Furthermore, a
possible population of stellar-mass black holes left over from the formation of Sgr~A$^\star$
itself \citep{Kroupa20} might mitigate the impact of the overall outward diffusion of the black
holes \citep{Generozov18} driven by their two-body encounters. Due to this complexity, we leave
a more detailed investigation of the long-term stability of the black hole cluster to future
work.
%
%--------------------------------------------------------------------------------------------------
%
\subsection{Dust-enshrouded objects}
\label{dusty}
Observations have revealed a population of dust-enshrouded objects near the outer boundary of the
S-cluster \citep{Ciurlo20, Peissker20, Peissker24}. Their formation scenario is still a matter
of an ongoing debate but they are believed to contain stellar cores of masses of a few solar masses.
During the grazing encounters of the black holes and stars considered here,
some fraction of the stripped stellar material can stay bound to the parent star (stellar core)
and condense to dust grains, leading to the formation of a dust-enshrouded object. We thus suggest
that these interactions may naturally explain the existence of the observed dust-enshrouded
objects.
%
%--------------------------------------------------------------------------------------------------
%
\subsection{Further considerations}
\label{further}
So far, we have been investigating the interaction of young stars and the stellar-mass black
holes on the timescale of a few dozen million years, considering the constraints
on the two populations derived from the observations of the current state of the Galactic centre.
In order to evaluate the interaction for older stars on longer timescales, the
stellar-dynamical evolution
of both populations, including the recurrent star formation episodes, would have to be taken into
account, which is beyond the scope of this letter.

Apart from physically affecting the stellar population, the rather numerous
stellar-mass black holes should also manifest themselves by
gravitational lensing of the individual stars within the S-cluster (or matter accreting onto
Sgr~A$^\star$ itself). Occasionally, this can lead to the emergence of transient stars that are
otherwise too faint to be detectable. Such stars could be searched for in the already existing
observational data.

Since the astrophysical conditions in the vicinity of Sgr~A$^\star$ can be regarded as rather generic
than exceptional among the supermassive black hole hosting galactic nuclei, the hypothesis
presented in this Letter is likely to be widely applicable. The analysis of the combined
gravitational waves signal coming from the dynamically evolving dense black hole clusters
in nearby galaxies could thus also provide useful insights into their evolution.
%
%--------------------------------------------------------------------------------------------------
%
\section{Conclusions}
In this Letter, we have investigated the impact of the direct collisions and grazing encounters
of the stellar-mass black holes and stars in the Galactic centre, considering three
qualitatively different formation channels for the black holes. Based on the obtained results,
we have constructed an order of magnitude radial density profile of the black hole cluster.

In particular, we have found that the spatial number density of the black holes can reach the
order of magnitude of $10^8$~pc$^{-3}$ within the outer parts of the S-cluster
(roughly at distances 0.01--0.04~pc from the
Sgr~A$^\star$) if they originate from the recurrent massive star formation in accretion
discs around Sgr~A$^\star$. The direct collisions (and grazing encounters) of such densely
distributed stellar-mass black holes with the individual stars lead to the depletion of the
most massive (and thus largest) stars on the timescale of a few million years.
Such a depletion can thus explain the reported absence of stars of stellar type O and
of the Galactic halo hypervelocity-star counterparts within the S-cluster.

Observationally confirmed stars within the S-cluster that are of spectral type B and a few million years old
rule out a black hole density of the order of magnitude of $10^9$~pc$^{-3}$ there. Similarly,
the abundant O stars farther than 0.04~pc from Sgr~A$^\star$
(i.e. beyond the outer boundary of the S-cluster) suggest a lower density of the stellar-mass
black holes in that region (a few times $10^7$~pc$^{-3}$ or less). Our results
further suggest a
bump-like shape of the black hole cluster radial density profile, with the inner decline
roughly within the orbital apocentre of the S2 star (closer than 0.01~pc to Sgr~A$^\star$).

\begin{acknowledgements}
We thank the anonymous referee for their useful comments that helped to improve this manuscript.
PK acknowledges support through the DAAD Eastern-European Bonn-Prague exchange programme.
MS is supported by the Grant Agency of Charles University under the grant no. 179123.
\end{acknowledgements}

\end{document}